# Lightwave-controlled band engineering in quantum materials

Sambit Mitra[1,2], Álvaro Jiménez-Galán[3,4]*, Marcel Neuhaus[1,2], Rui E F Silva[3,5], Volodymyr Pervak[1,2], Matthias F Kling[1,2,9,10] and Shubhadeep Biswas[1,2,9]*

[1] Max Planck Institute of Quantum Optics, Garching, Germany
[2] Physics Department, Ludwig-Maximilians-Universität Munich, Garching, Germany
[3] Max-Born-Institute, Berlin, Germany
[4] National Research Council of Canada, K1N 5A2, Ottawa, Canada
[5] Department of Theoretical Condensed Matter Physics, Universidad Autónoma de Madrid, Madrid, Spain.
[6] Technische Universität Berlin, Berlin, Germany
[9] SLAC National Accelerator Laboratory, Menlo Park, CA, USA.
[10] Department of Applied Physics, Stanford University, Stanford, CA, USA

**Stacking and twisting atom-thin sheets create superlattice structures with unique emergent properties, while tailored light fields can manipulate coherent electron transport on ultrafast timescales. The unification of these two approaches may lead to ultrafast creation and manipulation of band structure properties, which is a crucial objective for the advancement of quantum technology. Here, we address this by demonstrating a tailored lightwave-driven analogue to twisted layer stacking. This results in sub-femtosecond control of time-reversal symmetry breaking and thereby band structure engineering in a hexagonal boron nitride monolayer. The results practically demonstrate the realization of the topological Haldane model in an insulator. Twisting the lightwave relative to the lattice orientation enables switching between band configurations, providing unprecedented control over the magnitude and location of the band gap, and curvature. A resultant asymmetric population at complementary quantum valleys lead to a measurable valley Hall current, detected via optical harmonic polarimetry. The universality and robustness of the demonstrated sub-femtosecond control opens a new way to band structure engineering on the fly paving a way towards large-scale ultrafast quantum devices for real-world applications.**

Discrete symmetries in quantum mechanics, namely spatial inversion (SI) and time reversal (TR), play a pivotal role in governing the electronic and topological properties of materials[1-3]. For example, when both the symmetries are preserved in graphene, the highest valence and lowest conduction bands are connected by Dirac-like dispersion at the K and K' points of the Brillouin zone, enabling semimetal-like behaviour[2]. When the SI symmetry is broken, e.g. in monolayer hexagonal boron nitride (hBN) or transition metal dichalcogenides (TMDs), a bandgap ($\Delta$) appears along with a parabolic band dispersion around the K and K' points, leading to insulating (hBN: $\Delta_{K,K'} \approx 6.0$ eV) and semiconducting (TMDs: $\Delta_{K,K'} \approx 1.9 – 3.0$ eV) behaviour, respectively[1,4]. When the TR is broken, a topological phase transition to a Chern insulating state can be induced[1,3]. Control over these symmetries allows to engineer material properties, including those not found naturally in pristine materials. On the one hand, vertical stacking of 2D materials with matching spatial symmetry allows to create composite multi-layer structures with electronic properties that are different from those of the individual layers[5]. The ability to further manipulate these properties by adjusting the rotation angle between the layers provides a unique control handle, as exemplified by the topological[6] and superconductive[7] physics found at the magic angle of twisted bilayer graphene. On the other hand, the use of external time-dependent electric and magnetic fields allows to modify material properties transiently and reversibly. McIver et al.[8] recently reported the observation of a light-induced anomalous Hall effect in monolayer graphene using circularly polarized light fields. Here, the frequency, intensity and helicity of the external electric field are used as additional degrees of freedom to modify the Hamiltonian parameters of the crystal during the interaction. One of the key advantages of the use of external fields is that the laser-induced properties can be controlled in time, and can be thus used as switches. In Floquet engineering[9-12], the switching timescale is usually limited to the duration of the long pulse envelope. In lightwave-electronics[13-29], however, one exploits the control over the temporal characteristics of strong light fields, such as the carrier-envelope-phase of single-cycle pulses or the time-delay between multi-color fields, to manipulate coherent electronic transport on timescales shorter than one laser cycle.

---

*Corresponding authors. Email: shubha@slac.stanford.edu (S.B.); alvaro.jimenez@mbi-berlin.de (A.J.G.)

Here, we introduce a lightwave-driven counterpart to twisted layer stacking, where the composite material is the laser-dressed monolayer, and different properties are induced and manipulated on sub-cycle timescale by twisting the light field with respect to the material structure. We use the matching between the symmetry of the crystal lattice and the structured spatial waveform of the light field - a new degree of freedom that we control with sub-laser-cycle precision (see Fig. 1A). By rotating the structured lightwave in space, we demonstrate sub-cycle-controlled TR symmetry breaking and band structure engineering in an insulating hBN monolayer.

In the ground state, hexagonal 2D materials with broken SI symmetry possess local band minima at the K and K' high symmetry points, labelled as quantum valleys, which are energy-degenerate and separated by a large crystal momentum (Fig.1B). The K and K' valleys are related by time-reversal; when TR symmetry is broken, carriers at K and K' undergo different dynamics. One paradigmatic example is the toy model of a Chern insulator proposed by F. D. M. Haldane[1,3]. Here, TR-symmetry is envisioned broken by a staggered magnetic field in gapped graphene, inducing complex next-nearest neighbour hoppings that modify the relative band gap at the valleys. Another relevant example is given in the work of Xiao et al. [30], who showed that gapped graphene systems host a valley-dependent magnetic moment. This allows valley-selective excitation by resonant circularly-polarized light, which has been demonstrated in various transition metal dichalcogenides (TMDs)[31,32], developing the field of valleytronics[4,31-37]. Due to the resonant condition, valleytronics requires tuning of the central frequency of the light field to strong, material-dependent transitions, which usually involve excitonic states. Insulating materials such as hBN, with band gaps in the ultraviolet range, have remained out of reach for related applications, despite having favourable opto-mechanical properties, thermal conductivity, or chemical inertness[38].

In the following, we show that a polarization-tailored trefoil waveform induces complex next-nearest neighbour hoppings in the hBN insulating monolayer, which then becomes analogous to the topological model of Haldane[39]. The complex amplitude and phase of next-nearest neighbour hoppings is controlled by the spatial orientation of the light-waveform (see Fig. 1C), offering control over the band structure. As the trefoil waveform is rotated, the minimum band gap switches between the K and K' valleys (Fig. 1D-F). The light field that modifies the band structure is strong ($F_L \sim 1.0$ V/Å), with frequency well below the band gap ($\omega \ll \Delta_{K,K'}$), and thus additionally induces tunnel excitation. The latter maximizes at the minimum band gap valley, resulting in oscillating valley populations upon waveform rotation which are measured through time-delayed optical harmonic polarimetry [39]. Contrary to conventional valleytronics, our scheme does not rely on resonant processes for valley polarization; the effects are purely strong-field and symmetry-driven. It provides symmetry-protected, universal, and selective band engineering on the fly, a new paradigm in light-wave electronics and 2D material engineering.

We combine two counter-rotating circularly-polarized fields of frequencies ω and 2ω to produce a strong tailored lightwave whose projection on the hBN crystal plane resembles a trefoil structure, which matches the triangular lattice of hBN (see Fig.1C). For such a field, complex next-nearest neighbour (CNNN) hoppings are induced in the crystal through virtual nearest-neighbour hoppings[39] (see Fig. 1C and supplementary material (SM)). In analogy to the topological model of Haldane[3], the CNNN hoppings break TR symmetry and lift the degeneracy of the valleys, reducing the bandgap in one and increasing it in the other depending on the complex angle φ[39] (see Fig. 1C). The complex amplitude and angle of the laser-induced hoppings depend uniquely on the orientation of the vector potential, and not on its sense of rotation (see SM). This link between the spatial waveform of the vector potential and the band structure topology, is what allows to access the sub-laser-cycle timescale. Changing the sub-cycle phase delay between the ω and 2ω pulses rotates the vector potential, and modifies the CNNN hopping and band structure as shown in Fig. 2A-D. Due to the periodicity of both the trefoil waveform and the lattice, the same dynamics are repeated every 120°. Fig. 2E shows the predicted variation of the band gap at the K and K' valleys for different orientations of the vector potential relative to the first Brillouin zone of the hBN crystal. The modified band structure is calculated from the lowest-order time-dependent correction to the cycle-averaged Hamiltonian of a gapped graphene system, taking into account the experimentally used optical parameters (see SM).

Once the band gap at one of the valleys is lowered, the same strong trefoil field induces valley polarization through tunnel ionization. Fig. 2F-I shows the electron populations calculated after the interaction with a trefoil field for different spatial orientations. The time-dependent simulations were performed for a two-band model of gapped graphene using the crystal parameters of monolayer hBN. Electrons predominantly populate the valley where the band gap is reduced, leading to valley polarization. We note that there are two simultaneous mechanisms playing a role in the valley polarization. On the one hand, the trefoil-waveform-orientation dependent band gaps at K and K' valleys become asymmetric by the strong light field interaction, which is independent of the helicity of the vector potential. On the other, the valley-selective circular dichroism[30-32] based multi-photon excitation remains dependent on the helicity of the driving field uniquely. In our regime, the former largely dominates the valley excitation dynamics, as can be seen by the fact that: (i) the valley population oscillates as a function of the trefoil orientation, and (ii) for a fixed orientation of the vector potential, a change in helicity does not change the valley polarization, compare Fig. 2F-I and Fig. 2J-M. There are, however, small effects of the helicity-dependent valley selection rules apparent in the simulations. In particular, the electron populations at K and K' are not merely flipped between $\theta = 30°$ (Fig. 2G) and $\theta = 90°$ (Fig. 2I). For a complete flip between K and K', one must change both the orientation of the vector potential and helicity (cf. Fig. 2G and Fig. 2M, or Fig. 2I and Fig. 2K). Despite these residual helicity effects, the modulation and switch of the valley polarization as a function of waveform rotation angle is a tell-tale sign of the modification of the band structure topology.

We prepared the pump trefoil waveform by interferometrically combining two counter-rotating circularly polarized lightwaves of about 30 femtosecond (fs) long and with an amplitude ratio of 2:1 (see Fig. 3A). The broadband spectra of these components were centered around 2 μm (frequency: ω, photon energy: 0.6 eV) and 1 μm (frequency: 2ω, photon energy: 1.2 eV), respectively. We used a pump intensity of about 8 TW/cm$^2$ in air, which lies below the damage threshold of the 2D material, confirmed in our experiments (see SM). The 2ω light was generated through second harmonic generation of the ω light, so that they were mutually phase-stable. The orientation of the trefoil waveform with respect to the fixed hBN lattice structure was controlled by the sub-cycle time delay between the ω and 2ω lightwaves, which is altered through a relative optical path-length change. The interferometric stability of the trefoil waveform was ensured by another reference continuous wave (CW) laser interferometer which coexists with the ω-2ω pump interferometer. The hBN sample was grown by chemical vapour deposition method and transferred onto a 500 μm thick fused silica plate. To avoid averaging over multiple crystallographic orientations, we use microscopic, focused beams with spot size of about 10 μm, comparable to that of the single monocrystalline domains (see Fig. 3B). The quality, crystallographic orientation, and possible damage of the individual hBN monocrystalline domains were examined in-situ by separate polarization-resolved second-harmonic microscopy[40]. See SM for more details.

In order to read-out the valley polarization, we separated a small portion of the fundamental 2 μm (ω) beam, with linear polarization and pulse duration of 30 fs, and delayed it by about 100 fs with respect to the trefoil pump pulse. The linearly-polarized probe field generates a non-linear current along the polarization axis, as well as an anomalous (orthogonal) current that is proportional to the population asymmetry at the complimentary valleys[33,39,41,42]. The former is governed by the vector potential, whereas the latter is given by the vector product of the electric field and the Berry curvature. This leads to a 90° phase-delay difference between these components, and thus to the emission of elliptically-polarized harmonic radiation. As the Berry curvatures around the K and K' valleys exhibit opposite signs (see Fig. 3C), the anomalous component generated by electrons at K is completely out-of-phase (π) shifted with respect to those at K' (see Fig. 3D-F). As a result, the helicity of the outgoing harmonic ((third harmonic (3ω) in present case)) elliptical radiation maps the valley populations[39,43].

Experimentally, we determine both the helicity and ellipticity by detecting the s- and p-polarized components of the outgoing 3ω radiation by two photodiodes following a combination of quarter waveplate and Wollaston prism. During the experiment we rotated the pump trefoil waveform for a fixed probe pulse delay. The sample and thereby the lattice orientation was kept fixed. A lock-in detection scheme along with appropriate spectral filtering is used to capture the pump-induced signal 3ω modulations.

When the laser irradiates the fused silica only, we observed saturation of the 3ω signal for one of the diodes (high-background diode). This is due to a residual overlap of the bicircular pump and linearly-polarized probe pulses, which creates an unwanted wavemixing signal (see SM). When the helicities of the pump components are interchanged, this saturating signal switches to the other diode. This forced us to discard the signal from the high-background diode, and concentrate on the signal from the other (low-background) diode. In order to retrieve the full information of the valley polarization switching with one diode, we performed two complementary measurements for two opposing pump helicities.

The fused silica signal from the low-background diode (D1) is shown in Fig. 4A (red curve), and undergoes no modulation as a function of the pump trefoil rotation. In striking contrast, when the hBN sample is irradiated, D1 displays a clear oscillating signal (blue curve in Fig. 4A) with the expected $120^o$ periodicity. When the helicities of the pump components are interchanged, the low-background diode is now D2, which shows an oscillatory signal, again with a $120^o$ periodicity (green curve in Fig. 4A). Importantly, the signals from the two diodes are shifted by $60^o$, which reflects the switching of valley polarization every $60^o$ rotation of the trefoil waveform (see Fig. 2). A Fourier analysis, shown in Fig. 4B, confirms the dominance of the 120º periodic oscillations, which are absent when only the fused silica substrate is illuminated, and the $60^o$ out-of-phase oscillation between the two pump helicities (see Fig. 4C).

All these experimental features are consistent with the interpretation given above and are furthermore reproduced by time-dependent simulations on a gapped graphene system, as shown in Fig. 4D. As in the experiment, the simulations measure the 3ω signal for two opposing pump helicities (blue and green curves), corresponding to panels Fig. 2F-I and Fig. 2J-M, respectively. The 3ω signal is separated into its left and right circular components, which correspond to the signals measured at each of the diodes. For a fixed pump helicity configuration (blue solid and dashed curves), the left and right circular components of the 3ω signal have largely asymmetric amplitudes. This amplitude asymmetry reflects the fact that the population at one of the valleys oscillates much less than in the other, as we see in Fig. 2F-I or Fig. 2J-M, which is a consequence of the multi-photon valley-selective circular dichroism based contribution (see SM). The weakly oscillating component (dashed) falls into the high background diode, and is thus not resolved in the experiment. The situation reverses as we change the helicity of the pump (green solid and dashed curves). However, the outcomes of these complementary measurements are completely out-of-phase (π) shifted. The two largely oscillating signals (solid curves) are those observed in the experiment (Fig. 4C) and are a consequence of band structure modification by the rotating strong field of the tailored lightwave.

In summary, we have demonstrated a new route towards ultrafast material band engineering by controlling the shape and rotation of a lightwave in space, extending it to the sub-laser-cycle timescale and to the spatial domain. This is lightwave-driven counterpart to twisted layer stacking. The crystal-symmetry-matching, far-off-resonant strong light field induces a CNNN hopping that breaks TR symmetry in the laser-dressed hBN monolayer. In this way, we realize the light-analogue of the topological Haldane model in an insulating pristine material. The rotation of the lightwave controls the magnitude and location of the band gap, which enables non-resonant valley polarization switching in the hBN. The symmetry-driven band structure modification is robust, sub-femtosecond-controllable, reversible, and preserves electron coherence, which heralds ultrafast switching of quantum properties of quantum materials. The demonstrated lightwave-controlled material engineering paves the way towards real-world applications in quantum technology.

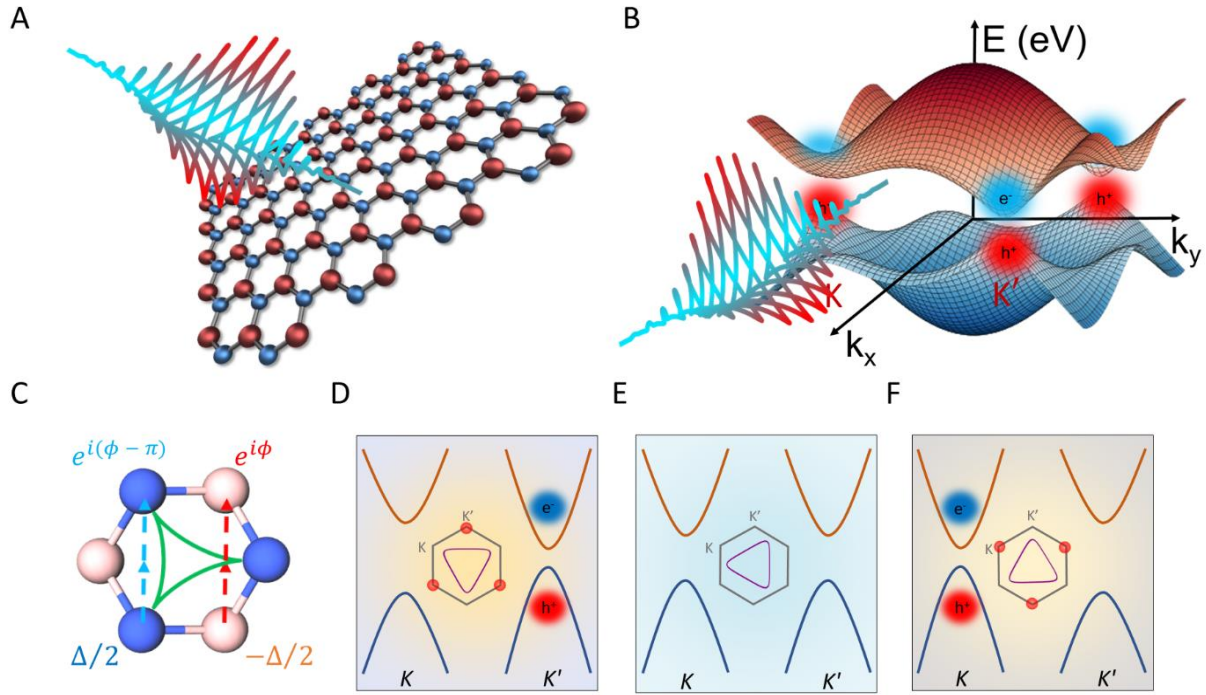

*Figure 1: Lightwave-controlled band structure engineering: (A) The tailored intense light waveform, resembling a trefoil structure on the lattice plane, is used to coherently manipulate the band structure of hBN. (B) For such 2d system the band structure shows hexagonal symmetry with energy minima (maxima), known as valleys, at K or K' (alternative to each other) points at conduction (valence) band. The valence and the conduction bands are separated by a direct bandgap ($\Delta_{K,K'}$: symmetric at K and K' points) which is of about 6.0 eV for hBN. Upon interaction with a trefoil waveform, the band structure gets modified by widening the bandgap at K (or K') valley and reducing it at K' (K) valley. This instigates electron excitation (from valence band to conduction band) asymmetry between K and K' valleys. (C) The band engineering is achieved by TR symmetry breaking, inducing a strong light field induced CNNN hopping interaction, indicated by the arrows reflecting Haldane's prescription[3]. The phase terms associated to these hopping interactions are indicated on the top. At the bottom the onsite energy difference of the atoms in the inversion symmetry broken system is indicated. (D - F) A sub-cycle-controlled rotation of the trefoil waveform relative to the hBN lattice orientation leads to a controlled band structure modification. Three schematic band structure configurations are shown which are resultant of interaction with three distinct trefoil vector potential waveform configurations. At (D) and (F), waveform maxima pointing either to K or K' valleys result in an asymmetry in the bandgaps ($\Delta_K \neq \Delta_{K'}$). When the waveform maxima point in between the two valleys, the band structure around K and K' remains symmetric ($\Delta_K = \Delta_{K'}$). The bandgap asymmetry is also reflected in resultant electron excitation.*

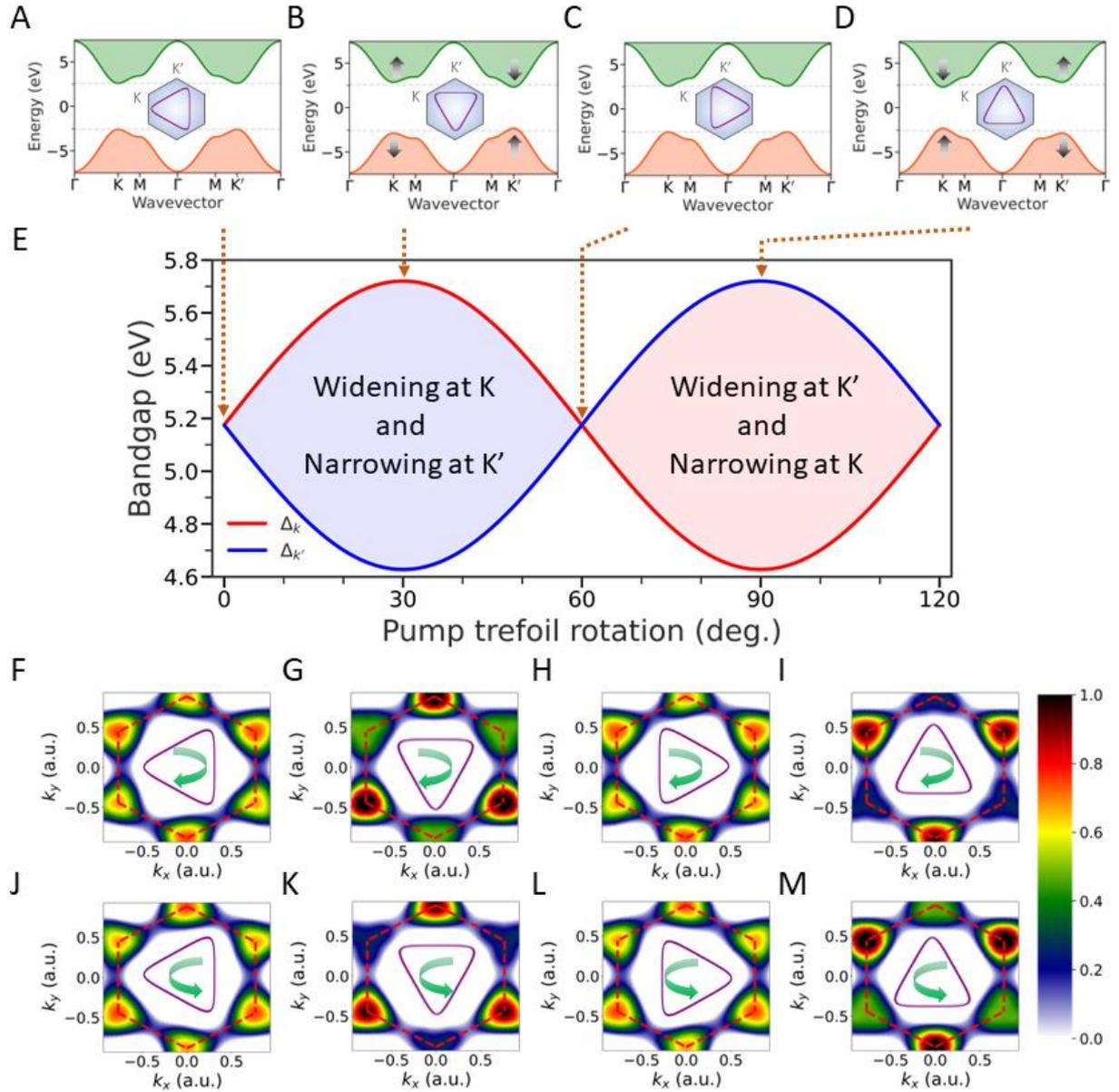

*Figure 2: Theoretical simulation of bicircular field interacting with gapped graphene:* (A-D) Band structure for different orientations of the bicircular vector potential (purple triangle inside the hexagonal Brillouin zone), (A) $\theta = 0º$, (B) $\theta = 30º$, (C) $\theta = 60º$, (D) $\theta = 90º$. The laser-induced nearest and complex next-nearest neighbour hoppings, used to calculate the band structure, were estimated from the lowest-order time-dependent perturbation to the cycle averaged Hamiltonian (see SM). (E) Band gap of the laser-dressed band structure at K and K' as a function of the orientation of the bicircular field $\theta$. (F-I) Normalized electron populations in the conduction band calculated using the code described in reference[44]. The red hexagon delimits the first Brillouin zone, with the K and K' points at the vertices, as indicated in the inset of panels (A-D). The purple triangle indicates the orientation of the vector potential, with the arrow showing the sense of rotation. (J-M) Same as the panels above (F-I), but for a vector potential rotating in the opposite direction.

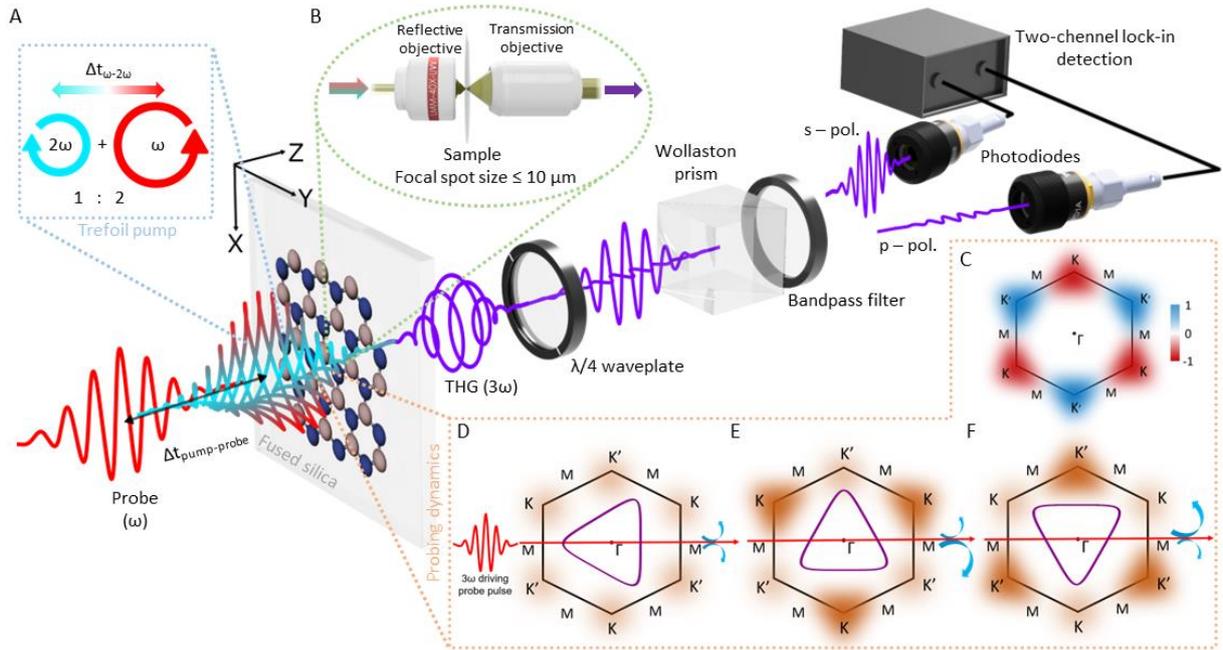

*Figure 3: All-optical methodology to control and probe band engineering with application to valleytronics:* (A) Monolayer hBN mounted on 500 µm thin fused silica substrate was pumped with an about 30 fs long tailored trefoil waveform. The lightwave was generated by interferometrically combining counter-rotating circularly polarized 2 µm ($\omega$) and 1 µm ($2\omega$) wavelength light with an amplitude ratio of 2 : 1. Its different orientation with respect to the real space lattice orientation is achieved by controlling the sub-cycle delay ($\Delta t_{\omega-2\omega}$) between $\omega$ and $2\omega$ pulses. The band structure and related electron dynamics were probed by optical harmonic polarimetry driven by a time delayed ($\Delta t_{pump-probe}$) linearly polarized 2 µm wavelength pulse. The polarization state of the generated third harmonic, analysed by quarter waveplate and Wollaston prism, encode the information about the induced valley polarization. Eventually, spectrally filtered and specially separated s and p polarized components are captured with photodiodes which are connected to two-channel lock-in amplifier for data acquisition. (B) The microscopic geometry was achieved by employing a reflective objective (NA: 0.4) which restricts the interaction region to be within 10 µm size, comparable to that of the mono-crystalline patch. The outgoing light was collected with a transmission objective (NA: 0.45). (C) The Berry curvature distribution of the conduction band for the monolayer hBN. (D - F) represent three representative probing configurations. In (D), K and K' valleys are equally populated resulting in a net zero analogous Hall current (depicted with curve blue arrows) causing a linearly polarized third harmonic. However, in (E) and (F), the asymmetry in electron population at the conduction band between K and K' valley results in a net non-zero Hall current in opposite directions. This results in induced ellipticity in third harmonic signal with opposite helicity for the above two cases.

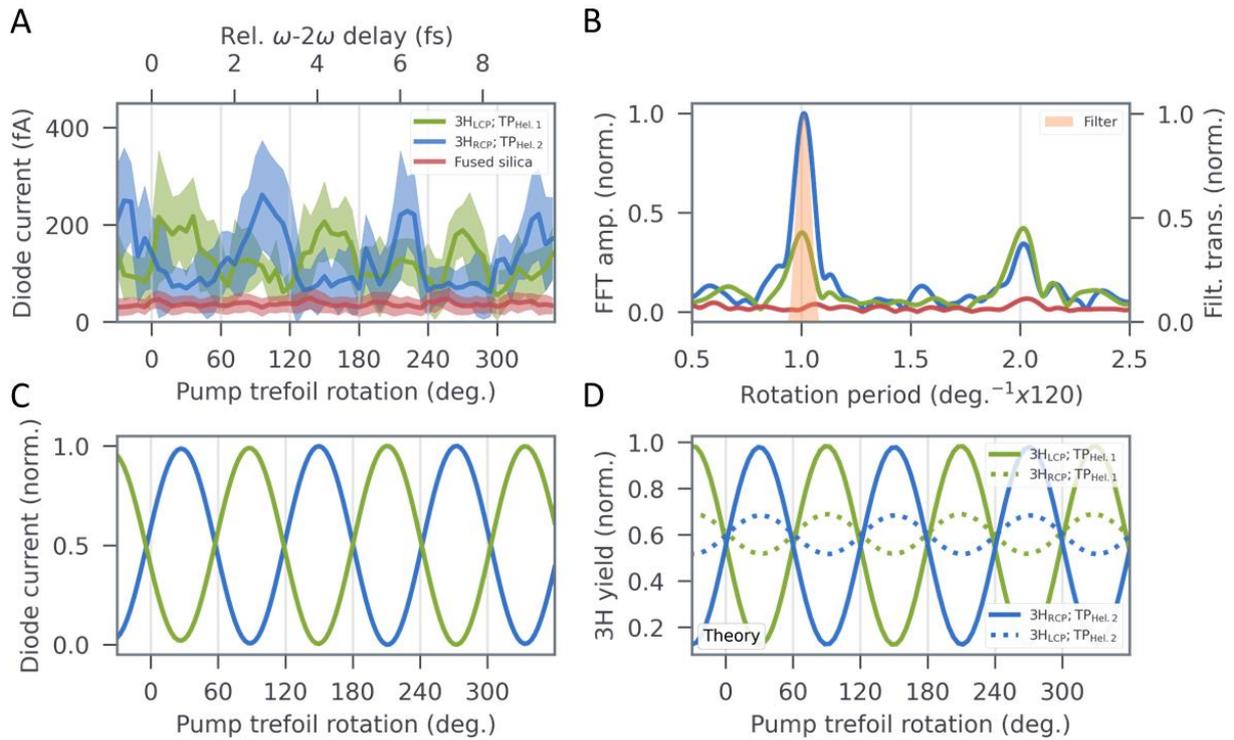

*Figure 4: Comparison between experimental and theoretical results:* Experimentally obtained helicity-resolved third harmonic (3H) from hBN as a function of the pump trefoil rotation. (A) Raw-experimental data of diode current from hBN for both helicities (Hel.) of trefoil pump (TP) and to that from fused silica. The solid line is produced by taking the statistical mean of all the photodiode-current data points recorded within a bin size of 3°, while the shaded region around depicts their respective standard deviation. The upper horizontal axis displays relative ω-2ω delay, which results in the rotation of the trefoil waveform. (B) FFT spectrum of the raw experimental data showing a distinct peak corresponding to 120° periodic oscillation, which is absent from the underlying substrate. (C) The inverse FFT of subplot (B) after applying a bandpass filter on the predominant spectral peak, clearly resolving an out of phase oscillation between the two different helicity configurations. (D) Calculated 3H yield as function of function of the pump trefoil rotation. The blue and green colours correspond to the opposite helicities of the pump trefoil, whereas the solid and the dashed lines correspond to the two opposite 3H helicities (expected detection by the two diodes) for a given pump helicity.

38    Molaei, M. J., Younas, M. & Rezakazemi, M. A Comprehensive Review on Recent Advances in Two-Dimensional (2D) Hexagonal Boron Nitride. *Acs Appl Electron Ma* **3**, 5165-5187 (2021).
39    Jimenez-Galan, A., Silva, R. E. F., Smirnova, O. & Ivanov, M. Lightwave control of topological properties in 2D materials for sub-cycle and non-resonant valley manipulation. *Nat Photonics* **14**, 728-732 (2020).
40    Li, Y. L. *et al.* Probing Symmetry Properties of Few-Layer MoS2 and h-BN by Optical Second-Harmonic Generation. *Nano Letters* **13**, 3329-3333 (2013).
41    Mak, K. F., McGill, K. L., Park, J. & McEuen, P. L. The valley Hall effect in MoS2 transistors. *Science* **344**, 1489-1492 (2014).
42    Silva, R. E. F., Jimenez-Galan, A., Amorim, B., Smirnova, O. & Ivanov, M. Topological strong-field physics on sub-laser-cycle timescale. *Nat Photonics* **13**, 849-854 (2019).
43    Mrudul, M. S., Jimenez-Galan, A., Ivanov, M. & Dixit, G. Light-induced valleytronics in pristine graphene. *Optica* **8**, 422-427 (2021).
44    Silva, R. E. F., Martin, F. & Ivanov, M. High harmonic generation in crystals using maximally localized Wannier functions. *Phys Rev B* **100**, 195201 (2019).
## ACKNOWLEDGEMENT

Á.J.G. acknowledges funding from the European Union's Horizon 2020 research and innovation programme under the Marie Skłodowska-Curie grant agreement no. 101028938. S.B. acknowledges support from the Alexander von Humboldt Foundation. S.B. and M.F.K.'s work at SLAC is supported by the U.S. Department of Energy, Office of Science, Basic Energy Sciences, Scientific User Facilities Division, and by the Chemical Sciences, Geosciences, and Biosciences division under award DE-SC0063. Fruitful discussions with Misha Ivanov and Olga Smirnova are gratefully acknowledged.

## AUTHOR CONTRIBUTIONS

S.B. initiated and designed the project in discussion with A.J.G., which was supervised by S.B. and M.F.K.. S.M. and S.B. built the experimental setup, performed the measurements, and analysed the data. S.M., A.J.G. and S.B. interpreted the results. A.J.G., and R.E.F.S. developed the theoretical model. M.N. supported laser operations. V.P. fabricated the specialized chirped and dielectric mirrors. A.J.G and S.B. wrote the manuscript with input from all co-authors.

## DATA AVAILABILITY

The data that support the findings of this study are available from the corresponding author upon reasonable request.

## CODE AVAILABILITY

The code used for the simulations contained in this study is available from the corresponding author upon reasonable request.

## COMPETING INTERESTS

The authors declare no competing interests.